\documentclass{Interspeech2024}
\usepackage{here}
\usepackage{multirow}
\usepackage{makecell}

\interspeechcameraready

\title{Efficient and Robust Long-Form Speech Recognition\\ with Hybrid H3-Conformer}

\name[affiliation={1}]{Tomoki}{Honda}
\name[affiliation={1}]{Shinsuke}{Sakai}
\name[affiliation={1}]{Tatsuya}{Kawahara}

\address{
  $^1$Kyoto University, Japan}
\email{honda.tomoki.34e@st.kyoto-u.ac.jp, sakai@sap.ist.i.kyoto-u.ac.jp, kawahara@i.kyoto-u.ac.jp}


\begin{document}

\maketitle

\begin{abstract}
    
    Recently, Conformer has achieved state-of-the-art performance in many speech recognition tasks. 
However, the Transformer-based models show significant deterioration for long-form speech, such as lectures, because the self-attention mechanism becomes unreliable with the computation of the square order of the input length. 
To solve the problem, we incorporate a kind of state-space model, Hungry Hungry Hippos (H3), to replace or complement the multi-head self-attention (MHSA). 
H3 allows for efficient modeling of long-form sequences with a linear-order computation. In experiments using two datasets of CSJ and LibriSpeech, our proposed H3-Conformer model performs efficient and robust recognition of long-form speech. 
Moreover, we propose a hybrid of H3 and MHSA and show that using H3 in higher layers and MHSA in lower layers provides significant improvement in online recognition. We also investigate a parallel use of H3 and MHSA in all layers, resulting in the best performance.

\end{abstract}

\section{Introduction}

In recent years, Transformer-based models have been widely used in many machine learning tasks. 
In automatic speech recognition (ASR), Conformer \cite{Conformer}, which incorporates a convolution layer after self-attention, has shown state-of-the-art performances in many tasks.
In these models, multi-head self-attention (MHSA) allows for a flexible mechanism for capturing important features with long-term dependency. 
As it involves a matrix operation for the combination of all input frames, its computation is quadratic in the input length, and training and inference become unreliable when the input becomes very long, resulting in a significant degradation in performance \cite{Conformer-long}.

This problem would be serious when dealing with long-form speech such as lectures and meetings. 
Thus, it is a common procedure to segment the input speech based on a long pause, but the segmentation becomes difficult in noisy conditions or inappropriate with irregular pauses due to disfluency. 
The problem would be critical in end-to-end speech translation, in which long-term dependency is critical, and the permutation of words can happen. Therefore, there is a demand for a sequence-to-sequence encoder-decoder framework that can handle long-form speech robustly and efficiently \cite{Long-Form-1,Long-Form-2}.

In this context, state-space models (SSMs) have been studied recently, mainly in the field of natural language processing (NLP), to achieve high performance with linear-order computation. 
Among them, the Structured State Space sequence model (S4) \cite{S4} has been shown to handle long-term dependency efficiently and has been introduced to the Transformer-based ASR decoder \cite{S4-Decoder}. 
More recently, Hungry Hungry Hippos (H3) \cite{H3} has been proposed as an extension of S4 and shown to achieve better performance in many NLP tasks. It realizes a mechanism similar to self-attention by incorporating SSM into the Linear-attention model.

In this paper, we present a long-form speech recognition model based on H3. As a naive implementation, H3 can be simply used to replace MHSA in Conformer. 
This model is named H3-Conformer. We also propose a novel model based on a hybrid of H3 and MHSA by selectively using either H3 or MHSA in Conformer encoder layers, which we call Hybrid H3-Conformer (CH4). 
Another variation of Parallel CH4, which uses both H3 and MHSA in parallel in each encoder layer, is also explored. 
In experimental evaluations using two datasets of the CSJ and LibriSpeech, we demonstrate that the proposed models outperform the conventional Conformer in online long-form ASR and that using H3 in higher layers is effective, suggesting that H3 is more capable of capturing long-term relationships.

\section{Background and Related Work}

\subsection{State Space Models}

     SSM is a model that handles long-term dependence efficiently and robustly by storing the history of time-series data based on a state-space representation, with HiPPO \cite{HiPPO} as a pioneer, LSSL \cite{LSSL}, S4 \cite{S4}, and other variants.
In the state-space representation, 
the following equation defines a mapping from an input sequence $\bm{u}=(u_1,...,u_L)\in\mathbb{R}^{L}$ to an output sequence $\bm{y}=(y_1,...,y_L)\in\mathbb{R}^{L}$ via an 
internal state vector $\bm{x}_t \in \mathbb{R}^{N}(0\le t \le L)$.
\begin{align} \label{eq:SSM_base}
    \begin{split}
        \bm{x}_t = \bm{A}\bm{x}_{t-1} + \bm{B}u_t\\
        y_t = \bm{C}\bm{x}_t + \bm{D}u_t
    \end{split}
\end{align}
where, $\bm{A}\in\mathbb{R}^{N \times N},\bm{B}\in\mathbb{R}^{N \times 1},\bm{C}\in\mathbb{R}^{1 \times N},\bm{D}\in\mathbb{R}^{1 \times 1}$.
By setting $\bm{x}_{0}=0$, the equation \eqref{eq:SSM_base} is expressed as equation \eqref{eq:SSM_y_simple}.
\begin{align} \label{eq:SSM_y_simple}
    \begin{split}
        y_k&=\bm{C}\bm{A}^{k-1}\bm{B}u_1+\cdots+\bm{C}\bm{B}u_k+\bm{D}u_k
    \end{split}
\end{align}
Let $\mathcal{K}_L$ be as follows.
\begin{align} \label{eq:Conv}
    \mathcal{K}_L(\bm{A},\bm{B},\bm{C})=(\bm{C}\bm{B},\bm{C}\bm{A}\bm{B},...,\bm{C}\bm{A}^{L-1}\bm{B})
\end{align}
Then, $\bm{y}$ can be expressed as a convolution.   
\begin{align} \label{eq:SSM_y}
    \bm{y}=\mathrm{SSM}(\bm{u})=\mathcal{K}_L(\bm{A},\bm{B},\bm{C})*\bm{u}+\bm{D}u_L
\end{align}
This eliminates the recursion and speeds up the computation process.
In S4, a representative SSM, the class of $\bm{A}$ is restricted to a class of matrices represented by the sum of a diagonal matrix and a low-rank matrix called a Diagonal Plus Low-Rank (DPLR) representation.
This restriction allows S4 to reduce the computational complexity of the convolution in the equation \eqref{eq:Conv} from $O(N^3L)$ to $O(N+L\log L)$.

S4 is also known for being the first to solve a task called PATH-X with accuracy better than random guessing in the Long-Range Arena \cite{LongRA}, 
a benchmark for uniformly evaluating performance in understanding and processing long-range dependencies.

\subsection{Speech recognition with State Space Models}

     As a previous study applying SSM to ASR, Miyazaki et al. \cite{S4-Decoder} introduced the S4 module into the decoder of the Conformer model.
It was shown to mitigate the degradation of the recognition accuracy for long-form ASR.
Meanwhile, Shan et al. \cite{S4-Encoder} introduced the S4 module into the encoder of the Conformer model, 
and improved the accuracy of ASR. However, long-form ASR was not investigated in this study.

In another study using SSM for ASR, Saon et al. \cite{DSS-ASR} proposed a model in which the depth-wise temporal convolutions in the Conformer architecture are replaced by Diagonal State Spaces (DSS) \cite{DSS}.

\subsection{Hungry Hungry Hippos (H3)}

     H3 is a model proposed by Fu et al. \cite{H3}, which uses SSMs as feature maps for a linear attention model \cite{Linear-Attention}, 
expecting to enhance long-term relationships while reducing computational complexity.

Let the length of the input be $L$ and the query/key/value tokens be $Q_i,K_i,V_i \in \mathbb{R}^{d}(1 \le i \le L)$.
The attention function in general, including softmax attention \cite{Attention} can be expressed as follows with the similarity function $Sim: \mathbb{R}^d \times \mathbb{R}^d \to \mathbb{R}$
\begin{align}\label{eq:Attention}
    O_i = \frac{\sum_{j=1}^{i} Sim(Q_j,K_j){V_j}}{\sum_{j=1}^{i} Sim(Q_j,K_j)}
\end{align}
Linear-attention assumes that Sim can be expressed as $Sim(q,k)=\phi(q)^T\phi(k)$ with a certain feature function $\phi$. Then the equation \eqref{eq:Attention} becomes
\begin{align}\label{eq:Linear-Attention}
    O_i = \frac{\phi(Q_i)^T \sum_{j=1}^{i} \phi(K_j){V_j}^T}{\phi(Q_i)^T \sum_{j=1}^{i} \phi(K_j)}
\end{align}
By defining $S_i = \sum_{j=1}^{i} \phi(K_j){V_j}^T$ and $z_i = \sum_{j=1}^{i} \phi(K_j)$, $O_i$ is expressed as follows.
\begin{align}\label{eq:Linear-Attention_replace}
    O_i = \frac{\phi(Q_i)^T S_i}{\phi(Q_i)^T z_i}
\end{align}
where $S_i$ and $z_i$ can be computed efficiently in advance by cumulative summing.
Linear-attention \cite{Linear-Attention} uses this, so to speak, “inverse kernel trick” to calculate attention efficiently. 
H3 incorporates SSMs into linear-attention by replacing $\phi(K_j)$ in the numerator of the equation \eqref{eq:Linear-Attention} with $SSM_{shift}$ 
and the sum $S_i$ with $SSM_{diag}$. It can be expressed by the following equation \cite{H3}.
\begin{align}
    \bm{Q}\odot\mathrm{SSM_{diag}}(\mathrm{SSM_{shift}}(\bm{K})\odot\bm{V})
\end{align}
where $\odot$ denotes the Hadamard product.
$\mathrm{SSM_{diag}}$ is an SSM that restricts matrix $\bm{A}\in\mathbb{R}^{N \times N}$ to a diagonal matrix.
It is the same as S4D by Gu et al. \cite{S4D}, and stores the state over the entire input sequence.
$\mathrm{SSM_{shift}}$ is an SSM that restricts matrix $\bm{A}$ as a shift matrix.
It constructs an internal state based on $\bm{A}\bm{x}_{t-1}$, 
by shifting each component of $\bm{x}_{t-1}$ according to equation \eqref{eq:SSM_base}.
From equation \eqref{eq:SSM_base}, it is apparent that H3 is a causal model referring to the past information only.
The computation of $\mathrm{SSM_{diag}}$ and $\mathrm{SSM_{shift}}$ can be performed with a computational complexity of $O(NL\log L)$, 
and the computational complexity of the entire H3 can be reduced to $O(N^2L+NL\log L)$.
While the computational complexity of general self-attention is $O(NL^2)$, 
H3 can suppress the increase in computational complexity to close to a proportion of the input length $L$.

H3 has been shown to perform better in several NLP tasks when combined with the attention layer \cite{H3}.
A kind of SSM called GSS has also been proposed to be combined with the attention layer by Mehta et al \cite{GSS}.

\section{Proposed models}

\subsection{H3-Conformer model}

\begin{figure}
    \centering
        \includegraphics[height=4cm]{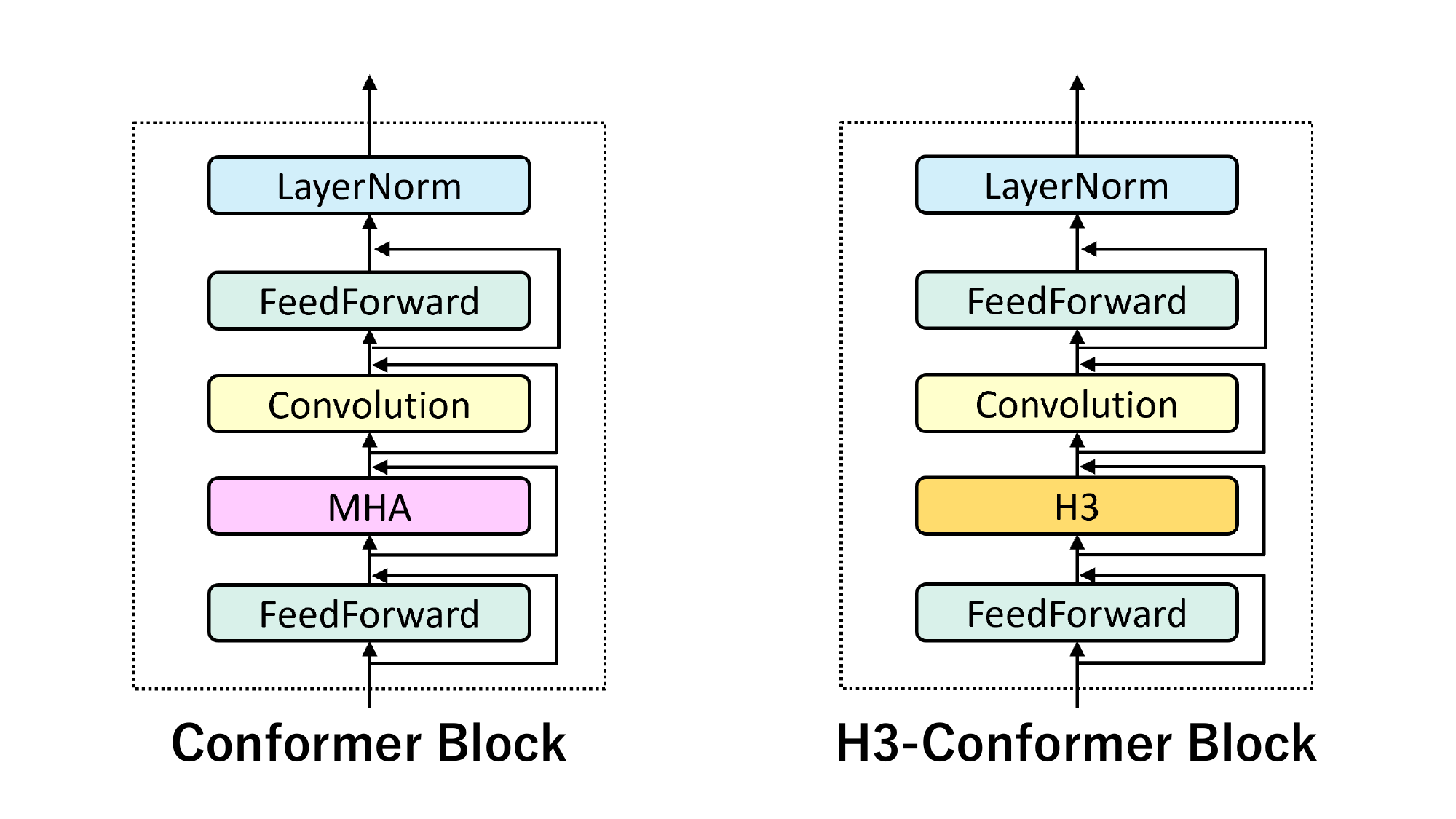}
    \caption{H3-Conformer}\label{Proposed models}
\end{figure}

     The Conformer block in the Confomer encoder consists of three modules: a feedforward layer, a multi-head self-attention (MHSA) layer, and a convolution layer (Figure \ref{Proposed models} left).
The convolution layer aggregates local features, while the MHSA layer is responsible for extracting global features.
The H3 layer is constructed based on linear attention and can process long-range dependencies. 
Thus, it can be used as a replacement for the MHSA layer of the Conformer block, and we name it H3-Conformer block (Figure \ref{Proposed models} right).
H3-Conformer model is defined by adopting H3-Conformer blocks in all encoder layers.

In the causal H3-Conformer for online ASR, the convolution layer is replaced by a causal convolution layer and the batch normalization layer is removed.

\subsection{Hybrid H3-Conformer(CH4) model}

     We also propose a hybrid model in which selected layers of the Conformer encoder are replaced by H3-Conformer blocks, 
which we call Hybrid H3-Conformer (CH4) model. We investigate on which layers H3 is more effective than MHSA in experiments.

\subsection{Parallel CH4 model}

\begin{figure}
    \centering
        \includegraphics[height=3cm]{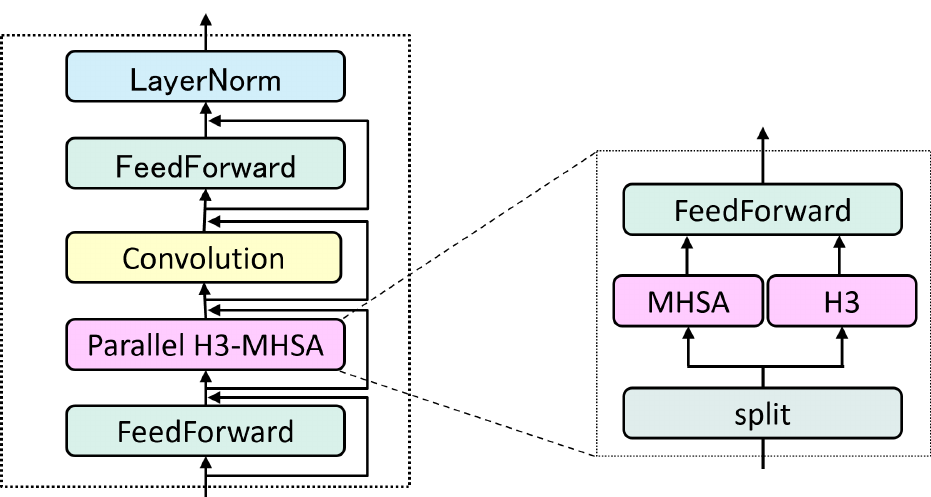}
    \caption{Parallel Hybrid H3-Conformer}\label{Parallel_block}
\end{figure}

     Furthermore, we explore a parallel use of the H3 layer and the MHSA layer. The input is divided into two parts, 
one is fed to the MHSA layer and the other to the H3 layer, and the combined output is given to the FeedForward layer for mixing. 
This module is defined as the Parallel H3-MHSA layer, shown in Figure \ref{Parallel_block}.
It is used in all layers, and we call it Parallel Hybrid H3-Conformer (Parallel CH4) model.


\section{Experimental evaluations}

\subsection{Experimental conditions}

     In the evaluation experiments, we used two datasets "Corpus of Spontaneous Japanese (CSJ)" \cite{CSJ} and LibriSpeech \cite{LibriSpeech} for training and evaluation.
We extracted 80-dimensional log-Mel filter-bank features with a frame shift of 10 ms and a frame size of 25 ms, 
and normalized them using the mean and variance of the entire training dataset. 
For the training data, speed perturbation\cite{SpeechPerturbation} and SpecAugment \cite{SpecAugment} were performed. In speed perturbation, the speech rate is transformed by a factor of 0.9 and 1.1.

As the output labels, CSJ uses Japanese characters with a vocabulary size of 3261.
For LibriSpeech, we performed byte-pair encoding \cite{bytepairencoding} tokenization with a vocabulary size of 1000.
To conduct long-form ASR evaluation, 24 consecutive utterances in an audio book or a lecture were extracted and concatenated to form a long-form input.
In all experiments, we use CTC \cite{CTC} because the focus of this study is the encoder of the Conformer.
In all models used in the experiments, the dimension of the encoder layers is 256, 
and there are 12 encoder layers in each model. 
The number of heads in the MHSA is fixed at 8 and that in the linear attention structure of the H3 layer is fixed at 2, 
except for the experiment to measure the processing time.
The optimization method was AdamW \cite{AdamW}, and 5\% of the training dataset was randomly selected and used as the validation dataset.
The maximum learning rate was always set to $1 \times 10^{-3}$, and learning rate decay was applied by cosine annealing.
Training was performed for 50 epochs in all experiments.

\subsection{Offline speech recognition}

\begin{table}[tb]

    \centering
    \caption{CER (\%) of \textbf{offline short-form} speech recognition on CSJ}
    \label{table:offline_csj}
    \resizebox{\linewidth}{!}{
    \begin{tabular}{cc|ccc}
    \hline
    model & size & eval1 & eval2 & eval3 \\
    \hline
    \hline
    Conformer & 23.7M & \textbf{6.08} & \textbf{4.45} & \textbf{4.92} \\
    H3-Conformer & 24.0M & 6.58 & 4.91 & 5.13  \\
    CH4 & 24.0M & 6.59 & 4.67 & 4.96 \\
    \hline
    \end{tabular}
    }
  
    \vspace{5mm} 

    \centering
    \caption{CER (\%) of \textbf{offline long-form} speech recognition on CSJ}
    \label{table:offline_csj_long}
    \resizebox{\linewidth}{!}{
    \begin{tabular}{cc|ccc}
    \hline
    model & size & eval1 & eval2 & eval3  \\
    \hline
    \hline
    Conformer & 23.7M  & 6.31 & 4.46 & 4.61 \\
    H3-Conformer & 24.0M  &6.16 & 4.31 & 4.54 \\
    CH4 & 24.0M  & \textbf{6.06} & \textbf{4.17} & \textbf{4.45} \\
    \hline
    \end{tabular}
    }
  
    \vspace{5mm} 
  
    \centering
    \caption{WER (\%) of \textbf{offline short-form} speech recognition on LibriSpeech}
    \label{table:offline_libri}
    \resizebox{\linewidth}{!}{
    \begin{tabular}{cc|cccc}
    \hline
    \multicolumn{2}{c|}{} & \multicolumn{2}{c}{dev} & \multicolumn{2}{c}{test} \\
    model & size  & clean & other & clean & other\\
    \hline
    \hline
    Conformer & 23.1M & \textbf{4.19} & \textbf{11.23} & \textbf{4.38} & \textbf{11.41}\\
    H3-Conformer & 23.4M & 5.09 & 14.03 & 5.28 & 14.18 \\
    CH4 & 23.4M & 5.16 & 13.93 & 5.37 & 14.17\\
    \hline
    \end{tabular}
    }

    \vspace{5mm} 
  
    \centering
    \caption{WER (\%) of \textbf{offline long-form} speech recognition on LibriSpeech}
    \label{table:offline_libri_long}
    \resizebox{\linewidth}{!}{
    \begin{tabular}{cc|cccc}
    \hline
    \multicolumn{2}{c|}{} & \multicolumn{2}{c}{dev} & \multicolumn{2}{c}{test} \\
    model & size  & clean & other & clean & other\\
    \hline
    \hline
    Conformer & 23.1M & 5.69 & 14.01 & 5.79 & 13.68 \\
    H3-Conformer & 23.4M & \textbf{4.86} & \textbf{13.12} & \textbf{5.15} & \textbf{13.26} \\
    CH4 & 23.4M & 5.22 & 14.27 & 5.47 & 14.33  \\
    \hline
    \end{tabular}
    }

  \end{table}

First, we conducted offline speech recognition.
Conventional Conformer, H3-Conformer, and CH4 were compared.
For the CH4 model, H3-Conformer blocks are used in the top 10 layers. 

The results of short-form and long-form ASR for CSJ and LibriSpeech are shown in Table \ref{table:offline_csj}, Table \ref{table:offline_csj_long}, Table \ref{table:offline_libri}, and Table \ref{table:offline_libri_long}, respectively. In the tables, ”size” refers to the number of trainable parameters.
In the short-form ASR, the baseline conventional Conformer performs the best in both datasets because Conformer uses both past and future information.
On the other hand, in the long-form ASR, CH4 was the best in CSJ, and H3-Conformer was the best in LibriSpeech, both significantly higher than the conventional Conformer at the 1\% level of significance.
The result confirms the robustness of H3 and CH4 in long-form ASR.

\subsection{Online speech recognition}

Next, we conducted online speech recognition using causal models.

\subsubsection{ASR results}

CH4 models with various settings of H3-Conformer blocks in different positions and numbers are compared using CSJ.
When the H3-Conformer block is used in all layers, the model is referred to as H3-Conformer, and CH4 uses H3-Conformer in some layers.
The results are shown in Table \ref{table:online_csj_long}.

In the online setting, overall accuracy is degraded from the offline setting (Table \ref{table:offline_csj_long}).
The degradation is larger for the conventional Conformer, suggesting that SSMs perform better with online ASR.
S4-Conformer and H3-Conformer showed significantly better performance than the conventional Conformer at the 1\% significance level.
The CH4 model with H3-Conformer blocks in the top 10 layers achieved even better accuracy compared to the H3-Conformer using H3 in all 12 layers, 
and this difference was significant at the 1\% significance level.
It should be noted that even when the number of H3-Conformer blocks is the same, models using H3 in higher layers tend to perform better than those using it in bottom layers.
This difference is significant at the 1\% significance level for the three and six layer cases.
This shows that the H3 layer performs better than the MHSA layer for global processing in higher layers.

We also evaluated the online long-form ASR performance with LibriSpeech.
Based on the results from CSJ, we used a model with the H3-Conformer blocks in the top 10 layers as causal CH4.
The results are shown in Table \ref{table:online_libri_long}. The performance of CH4 and H3-Conformer significantly outperforms the baseline causal Conformer.
The difference between Conformer and CH4 and that between Conformer and H3-Conformer were significant at the 1\% significance level.
On the other hand, there were no significant performance differences between CH4 and H3-Conformer.

\begin{table}[tb]


    \centering
    \caption{CER (\%) of \textbf{online long-form} speech recognition on CSJ}
    \label{table:online_csj_long}
    \resizebox{\linewidth}{!}{
    \begin{tabular}{clc|ccc}
    \hline
     \multicolumn{2}{c}{model}  & size & eval1 & eval2 & eval3 \\
    \hline
    \hline
    \multicolumn{2}{c}{Conformer} &  23.7M &  10.20 & 8.07 & 8.62 \\
    \multicolumn{2}{c}{\makecell{S4-Conformer \\ (S4 in all layers)}} & 22.5M  & 9.64 & 6.96 & 7.48 \\
    \multicolumn{2}{c}{\makecell{H3-Conformer \\ (H3 in all layers)}} & 24.0M  & 9.50 & 6.74 & 7.20 \\
    \hline
    \multirow{6}{*}{CH4} & H3 in bottom 3 layers & 23.8M& 10.30 & 8.12 & 8.81 \\
    & H3 in top 3 layers & 23.8M & 9.77      & 7.21      & 8.04 \\
    & H3 in bottom 6 layers& 23.8M &  10.18     & 8.07      & 8.33 \\
    & H3 in top 6 layers& 23.8M & 9.35      & 6.82      & 7.41 \\
    & H3 in bottom 10 layers& 24.0M  & 9.06      & 6.79      & 7.10 \\
    & H3 in top 10 layers& 24.0M    & \textbf{9.04} & \textbf{6.44} & \textbf{6.99} \\
    \hline
    \end{tabular}
        }
\end{table}

\begin{table}[tb]


    \centering
    \caption{WER (\%) of \textbf{online long-form} speech recognition on LibriSpeech}
    \label{table:online_libri_long}
    \resizebox{\linewidth}{!}{
    \begin{tabular}{cc|cccc}
    \hline
     & & \multicolumn{2}{c}{dev} & \multicolumn{2}{c}{test} \\
    model & size & clean & other & clean & other \\
    \hline
    \hline
    Conformer & 23.7M & 12.30 & 24.04 & 12.89 & 24.49 \\
    H3-Conformer & 24.0M & 7.86 & 18.82 & 8.17 & 19.12 \\
    CH4 & 24.0M & \textbf{7.64} & \textbf{18.79} & \textbf{8.10} & \textbf{19.11} \\
    \hline
    \end{tabular}
    }
\end{table}

\subsubsection{Parallel CH4 model}

Using CSJ, we trained causal Parallel CH4 models with various settings for the number of input/output dimensions in the MHSA and the H3 layers.
The long-form ASR results for these models are shown in Table \ref{table:parallel_long}.
Note that the model with a 0-dim H3 layer is equivalent to the conventional Conformer and the model with a 0-dim MHSA layer is equivalent to the H3-Conformer.

The parallel model with a larger proportion of H3 layer dimensions tends to perform better. 
The differences in performance of the model with 224 dimensions in the H3 layer compared to the other four models in the table are all significant at the 1\% level of significance.

We looked at the absolute values of the learned weights of the first linear layer in the FeedForward layer of the Parallel H3-MHSA block and found that the output of the MHSA layer is mostly used in lower layers, and the output of the H3 layer is preferentially used in the higher layers.
This result indicates that the MHSA layer is more effective than the H3 layer in the lower layers, while the H3 layer is more effective in higher layers.
This difference in characteristics between the H3 layer and the MHSA layer may be the reason why the hybrid model with the MHSA in the lower layers and the H3 in the higher layers performed better than other models.


\begin{table}[t]


    \centering
    \caption{CER(\%) by causal Parallel CH4 for \textbf{online long-form} speech recognition on CSJ}
    \label{table:parallel_long}
    \resizebox{\linewidth}{!}{
    \begin{tabular}{ccc|ccc}
    \hline
    \multicolumn{3}{c|}{model} & \\
    MHSA layer & H3 layer & size & eval1 & eval2 & eval3 \\
    \hline
    \hline
    0 dim& 256 dim & 24.0M & 9.50 & 6.74 & 7.20  \\
    32 dim& 224 dim& 30.3M  & \textbf{8.61} & \textbf{6.34} & \textbf{6.68}  \\
    128 dim& 128 dim& 29.6M& 10.96 & 8.75 & 8.75  \\
    224 dim& 32 dim& 29.7M & 13.67 & 11.48 & 11.10 \\
    256 dim& 0 dim& 23.7M & 10.20 & 8.07 & 8.62 \\
    \hline
    \end{tabular}
    }
\end{table}

\subsubsection{Processing speed in long-form speech recognition}

Processing times are compared among causal Conformer, causal CH4, and causal H3-Conformer using LibriSpeech.
Here, we changed the size of long-form speech by changing the number of utterances concatenated in the test set.
For each number of concatenated utterances, we calculated the real-time factor (RTF), which is the ratio of the total processing time to the total length of the input speech, using all the data in the test-other dataset.

\begin{figure}
    \centering
        \includegraphics[width=\linewidth]{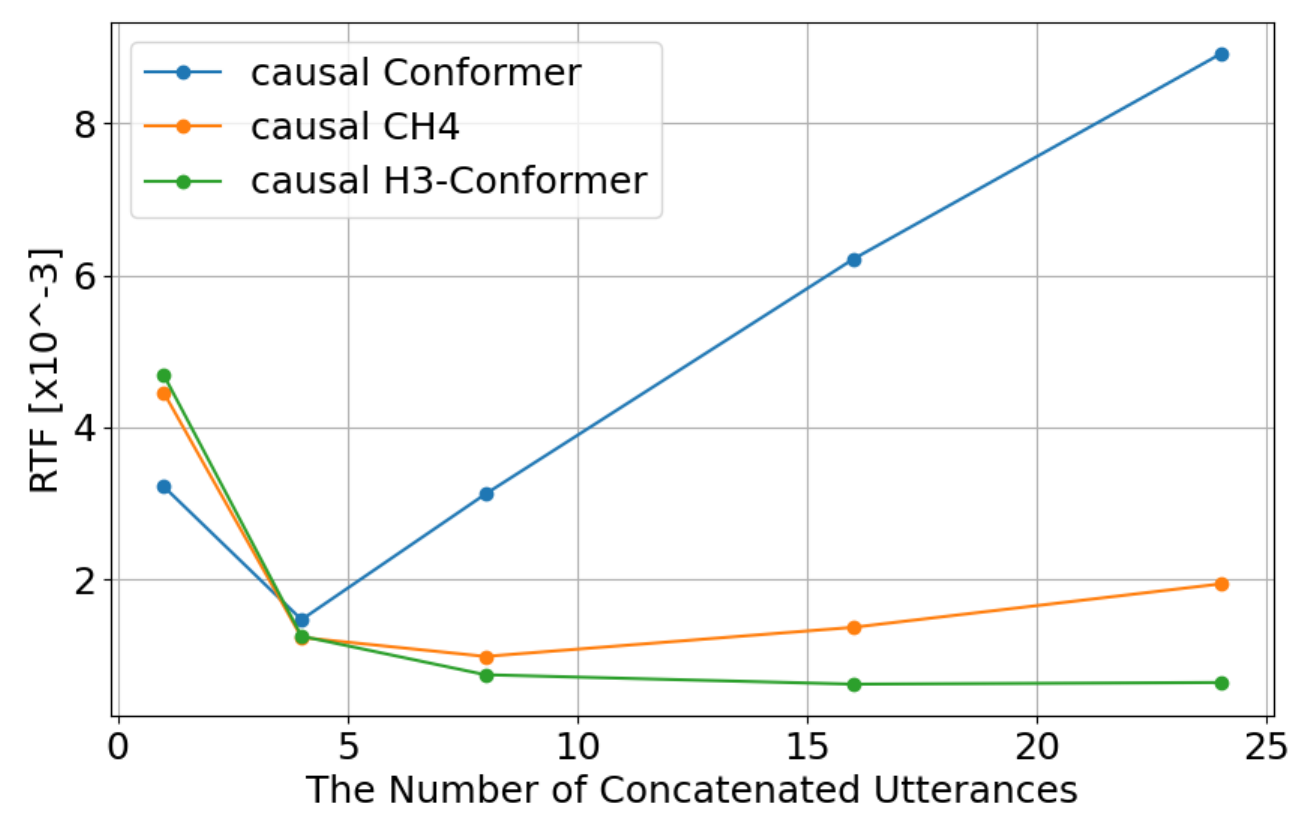}
    \caption{Processing time for long-form speech recognition}\label{speed}
\end{figure}

One Titan RTX with 16.3 TFlops and memory of 24 GB was used as a GPU for the experiments.
For fairness, the number of heads of the linear attention structure in the H3 layer was set to 8, the same number of heads in the MHSA layer. 
The results are shown in Figure \ref{speed}. Note that the values shown in the figure are the average of five measurements. 
In the conventional Conformer, the RTF increases in proportion to the number of concatenated utterances, while the RTF does not increase in H3-Conformer and slightly increases in CH4.
When the number of concatenated utterances is very small, the RTF gets large because of the effect of overhead time.

\section{Conclusion}

In this study, we have proposed a CH4 model that incorporates the Hungry Hungry Hippos (H3) layer, a type of SSM, into the Conformer model to 
improve the robustness of online long-form ASR. The model was evaluated using datasets of both Japanese and English.
The effectiveness of the H3 layer was clearly observed in the online long-form ASR setting. Experimental results suggest that the H3 layer can handle long input lengths unseen in the training time more robustly compared to the MHSA layer.
It was shown that further improvement is obtained by using both the H3 layers and the MHSA layers,
and the H3 layer was better than MHSA for processing global features in higher layers.
%

Future research directions include applying this model to end-to-end speech translation tasks.





\bibliographystyle{IEEEtran}
\bibliography{mybib}

\end{document}